\documentclass[aps,pra,twocolumn,superscriptaddress,10pt,showpacs]{revtex4-1}

\usepackage{graphicx,hyperref}
\newcommand{\Eq}[2][Eq.~]{#1(\ref{eq:#2})}

\newcommand{\I}{\mathrm{i}}
\newcommand{\Exp}[1]{\mathrm{e}^{\mbox{\footnotesize$#1$}}}
\newcommand{\ket}[1]{{\left|{#1}\right\rangle}}
\newcommand{\KET}[4]{\ket{#1;\textsf{#2},\textsf{#3},\textsf{#4}}}

\begin{document}

\title{Reply to ``Comment on `Past of a quantum particle revisited'\,''}

\author{Berthold-Georg Englert}\email{cqtebg@nus.edu.sg}
\affiliation{Centre for Quantum Technologies, National University of %
  Singapore, 3 Science Drive 2, Singapore 117543, Singapore} 
\affiliation{Department of Physics, National University of Singapore, %
  2 Science Drive 3, Singapore 117542, Singapore} 
\affiliation{MajuLab, CNRS-UNS-NUS-NTU International Joint Unit, %
  UMI 3654, Singapore} 

\author{Kelvin Horia}\email{e0383775@u.nus.edu}%
\affiliation{Department of Mathematics, National University of Singapore, %
  10 Lower Kent Ridge Road, Singapore 119076 }

\author{Jibo Dai}\email{dai\_jibo@imre.a-star.edu.sg}
\affiliation{Institute of Materials Research and Engineering, %
             Agency for Science, Technology and Research, %
             2 Fusionopolis Way, \#08-03 Innovis, %
             Singapore 138634, Singapore}

\author{Yink Loong Len}\email{yinkloong@quantumlah.org}
\affiliation{Centre for Quantum Technologies, National University of %
  Singapore, 3 Science Drive 2, Singapore 117543, Singapore} 

\author{Hui Khoon Ng}\email{cqtnhk@nus.edu.sg}
\affiliation{Yale-NUS College, 16 College Avenue West, Singapore 138527, %
  Singapore} 
\affiliation{Centre for Quantum Technologies, National University of %
  Singapore, 3 Science Drive 2, Singapore 117543, Singapore} 
\affiliation{MajuLab, CNRS-UNS-NUS-NTU International Joint Unit, %
  UMI 3654, Singapore} 

\date[]{Posted on the arXiv on 17 January 2019}

\begin{abstract}
  We stand by our findings in \pra\ \textbf{96}, 022126 (2017).
  In addition to refuting the invalid objections raised by Peleg and Vaidman,
  we report a retrocausation problem inherent in Vaidman's definition of the
  past of a quantum particle. 
\end{abstract}

\maketitle

In their Comment \cite{Peleg+1:18} on Ref.~\cite{Englert+4:17},
Peleg and Vaidman correctly state that there is a fundamental difference in
the way Vaidman answers the question \textit{Where was the particle after
  entering and before leaving the interferometer?\/} and our traditional
approach.
We examine the traces left by the particle --- by an unambiguous path
discrimination measurement or by any other suitable method --- and then infer
the path in accordance with the result found.
This can yield definite path knowledge, or probabilistic path knowledge, or no
path knowledge at all, depending on how the traces are examined and what is
found.
For us, then, a statement such as ``this particle went through checkpoint C''
has an operational meaning, which derives from interpreting measurement
results by the usual applications of the tools of standard quantum mechanics.

By contrast, Vaidman \emph{defines} that the particle was where it left faint
traces --- meaning: in \emph{all} places where computed weak values are
nonzero, for which the particle may have to be in several places
simultaneously.
Even when the traces are indiscernible and of no phenomenological consequence,
Vaidman maintains that the particle has visited all places for which the
computation yields a nonzero weak value.
In the context of Vaidman's three-path interferometer, these are all places
where both wave functions are nonzero, namely the usual ``forward'' wave
function that emerges from the source and the ``backward'' wave function that
interferes constructively at the actual output port only.

\begin{figure}
  \centering
  \includegraphics{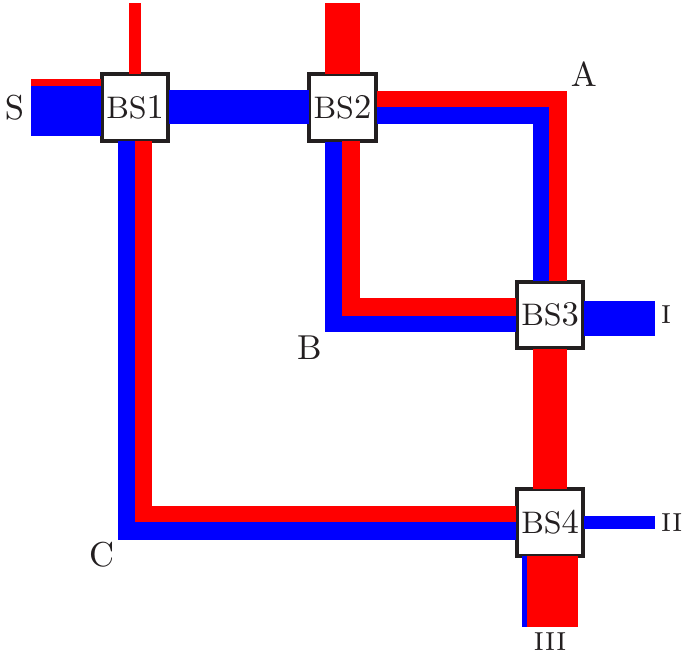}
  \caption{\label{fig:reply1}%
    The source S emits particles into the three-path interferometer.
    Beam splitters BS2 and BS3 are symmetric and the inner-loop
    interferometer is balanced.
    Beam splitters BS1 and BS4 are asymmetric, they transmit two-thirds of the
    incident particles and reflect one-third.
    Only particles detected at exit \textsc{iii} are considered.
    The blue lines follow the usual ``forward'' wave function that emerges
    from S, with the line thickness proportional to the probabilities (squared
    amplitudes). 
    The red lines follow the ``backward'' wave function that has unit
    probability for delivering the particle to exit \textsc{iii}, with the
    same significance of the line thickness.
    While both wave functions have nonzero amplitudes in the inner loop, the
    only blue or red links from the source S to the exit \textsc{iii} bypass
    the inner loop and go through checkpoint C.}
\end{figure}

The situation is depicted in Fig.~\ref{fig:reply1}, for a particle detected at
exit \textsc{iii}, a variant of Fig.~2 in \cite{Englert+4:17}.
The thickness of the blue lines is proportional to the squared amplitudes of
the forward wave function, and likewise for the red lines and the backward
wave function.
This picture refers to the extreme situation of indiscernible traces
(${\epsilon=0}$ in \cite{Englert+4:17}), or to the conditioning on the
inconclusive outcome of the unambiguous path discrimination.

We observe that the wave function branches associated with the checkpoints A
and B inside the inner loop, whether blue or red,
do not connect the source S to exit \textsc{iii}.
According to standard quantum mechanics, therefore, these branches are
irrelevant as they do not contribute to the probability of detecting the
particle at this exit. 
The only link is through the checkpoint~C.
This is what standard quantum mechanics says about preselected ($\equiv$
emitted by source S) and postselected ($\equiv$ detected at exit \textsc{iii})
particles.
Accordingly, Peleg and Vaidman's assertion that ``the standard formalism has no
any answer'' for pre- and postselected particles is incorrect.

According to Vaidman's definition, however, each particle detected at exit
\textsc{iii} left a faint trace at checkpoint~A and at checkpoint B and also
at checkpoint C on its way from the source to the detector; each particle was
at all three checkpoints simultaneously.
Peleg and Vaidman tell us that one must not examine the traces of a single
particle (as one would usually do before making statements about the
particle's whereabouts) but that the ``information about the traces is
obtained either by a calculation, or by a measurement performed on the pre-
and postselected ensemble.''
Nevertheless, Vaidman insists that each particle individually was at the three
checkpoints simultaneously --- and the evidence are the faint traces that the
particle left at each checkpoint but they must not be examined~\dots

Central to Vaidman's reasoning is that the particle interacts weakly with the
stuff it encounters on its way, such as transferring a bit of momentum to the
mirror at a checkpoint.
In a simplified description~\cite{fn1}, the particle's passing
through the interferometer results in the before-to-after transition
\begin{eqnarray}
  \label{eq:0}
  \KET{(\textsf{in})}{no}{no}{no}&\to&\frac{1}{\sqrt{3}}\Bigl(
  \KET{(\textsf{out})_{\textsc{a}}^{\ }}{yes}{no}{no}
  \nonumber\\&&\mbox{}+
  \KET{(\textsf{out})_{\textsc{b}}^{\ }}{no}{yes}{no}
  \nonumber\\&&\mbox{}+
  \KET{(\textsf{out})_{\textsc{c}}^{\ }}{no}{no}{yes}\Bigr)\,,
\end{eqnarray}
where $(\textsf{in})$ and $(\textsf{out})$ symbolize sets of quantum numbers
for the particle's center-of-mass motion and, for example,
$\KET{(\textsf{out})_{\textsc{a}}}{yes}{no}{no}$
stands for a trace at checkpoint A and no traces at checkpoints B and C.
An individual particle leaves a trace at one of the checkpoints.
Since there are ``yes'' terms for all three checkpoints, a measurement on a
large ensemble of particles would exhibit evidence for the traces of all three
kinds.
In Vaidman's reading, however, \Eq{0} states that \emph{each} particle leaves
traces at all checkpoints although there is no
$\KET{(\textsf{out})}{yes}{yes}{yes}$ term to account for that~\cite{fn2}.

Despite being told otherwise, we did examine the traces of a single particle
in Ref.~\cite{Englert+4:17} and found that all particles in the pre- and
postselected ensemble went only through checkpoint C in the limit of
ultrafaint traces (${\epsilon\to0}$).
This is exactly what standard quantum mechanics tells us (see above).
It is also what common sense tells us in conjunction with basic knowledge
about interferometers.

In standard quantum mechanics, there are situations in which questions such as
\textit{Through which arm of the interferometer did the particle arrive?} do
not have an answer, and then we insist that \textit{We do not know.} is
the correct reply. 
The basic example is that of a well stabilized two-path interferometer, say of
Mach--Zehnder design.
Whenever standard quantum mechanics does not provide an answer, one may feel
invited to \emph{define} the past of the particle in a fitting way.
Vaidman accepted this invitation, as advocates of Bohmian mechanics had done
earlier.

Now, while providing answers where there were none before, such definitions
must always give the correct answer in all situations in which there already
is one without the added definition.
This is a basic test of consistency.
Bohmian mechanics fails this test although it took some time before an example
was found that demonstrates the case \cite{Englert+3:92,Aharonov+1:96}.
Vaidman's definition fails the test, too, as we established by our analysis of
the three-path interferometer that he himself designed.

Peleg and Vaidman disagree with this verdict.
They claim that our ``argument for a particular single-path story can be
repeated equally well for another single path'' and if that were true it would
indeed imply that we contradict ourselves.
But below we show it is not true.

In our analysis of the three-path interferometer of Fig.~\ref{fig:reply1},
with a balanced inner-loop interferometer, we observe that the probability of
detecting the particle at exit \textsc{iii} does not change when we introduce
a phase $\gamma$ into the amplitude at checkpoint~C.
Since this is a relative phase between the amplitudes that meet at beam
splitter BS4, and are then processed by BS4, we conclude that these phases are
incoherent.
Upon this observation, we then proceed with the accounting exercise in
Sec.~IV\,B of \cite{Englert+4:17} that culminates in the conclusion that all
particles reach exit \textsc{iii} through checkpoint~C when ${\epsilon\to0}$.

Peleg and Vaidman did not find an error in this argument.
Instead they try to construct a contradiction --- and fail.
They begin by recalling that the said probability is~\cite{fn3}
\begin{equation}
  \label{eq:1}
  p(\alpha,\beta,\gamma)=\epsilon
  +\frac{1-3\epsilon}{9}
  \Bigl|\Exp{\I\gamma}+\Exp{\I\beta}-\Exp{\I\alpha}\Bigr|^2\,,
\end{equation}
when phase factors $\Exp{\I\alpha}$, $\Exp{\I\beta}$, and $\Exp{\I\gamma}$
multiply the probability amplitudes at checkpoints A, B, and C, respectively.
Yes, as noted above, there is no dependence on $\gamma$ when the inner-loop
interferometer is balanced (${\Exp{\I\alpha}=\Exp{\I\beta}}$), and so we get
our ``argument for a particular single-path story'' --- via~C, that is.
Peleg and Vaidman then consider the situation of
${\Exp{\I\alpha}=\Exp{\I\gamma}}$ and note that the dependence on $\beta$
disappears, and so they wrongly conclude that there is equally strong
evidence for the ``via~B'' single-path story.

Why is this conclusion wrong?
Simply because our accounting exercise cannot be carried out when the balance
of the inner-loop interferometer is disturbed
(${\Exp{\I\alpha}\neq\Exp{\I\beta}}$); there are now coherences between the
amplitudes arriving at BS4.

This becomes obvious if we replace BS4 with a beam splitter with more general
properties, one that reflects with probability $R$ and transmits with
probability ${T=1-R}$.
Then \Eq{1} is replaced by
\begin{equation}
  \label{eq:2}
  p(\alpha,\beta,\gamma)=\epsilon
  +\frac{1-3\epsilon}{3}\Biggl|\Exp{\I\gamma}\sqrt{R}
  +\Bigl(\Exp{\I\beta}-\Exp{\I\alpha}\Bigr)\sqrt{\frac{T}{2}}\Biggr|^2\,,
\end{equation}
and we get
\begin{equation}
  \label{eq:3}
  p(\alpha,\alpha,\gamma)=\epsilon+\frac{1-3\epsilon}{3}R
\end{equation}
when the inner-loop interferometer is balanced.
Figure~\ref{fig:reply1} applies for all intermediate $R$ values
(${0<R<1}$) with adjusted line thicknesses, not only for ${R=\frac{1}{3}}$.
While there is no $\gamma$ dependence in \Eq{3}, confirming that incoherent
amplitudes meet at BS4, there is a dependence on ${\alpha-\beta}$, the
relative phase of the inner-loop interferometer, in
\begin{eqnarray}
  \label{eq:4}
  p(\alpha,\beta,\alpha)&=&\epsilon+\frac{1-3\epsilon}{3}\Biggl[1-\sqrt{2RT}
                            \nonumber\\&&
 \rule{3em}{0ex}\mbox{}
   +\bigl(\sqrt{2RT}-T\bigr)\cos(\alpha-\beta)\Biggr]\,,\quad
\end{eqnarray}
and this dependence disappears only when $R=\frac{1}{3},T=\frac{2}{3}$,
the very particular case of \Eq{1} (or when ${R=1}$, see below).

It follows that the accounting exercise is justified for ${\alpha=\beta}$,
when it leads to the ``all via~C'' conclusion, but it is not justified for
${\alpha=\gamma\neq\beta}$.
The reasoning put forward by Peleg and Vaidman, who wrongly conclude that
``all via~B'' is as valid as ``all via~C,'' is of no consequence.

Other objections raised by Peleg and Vaidman are equally invalid.
It is not necessary that we address them all.
Instead, we point out that Vaidman's definition has yet another implication
that speaks against adopting it. 

\begin{figure}
  \centering
  \includegraphics{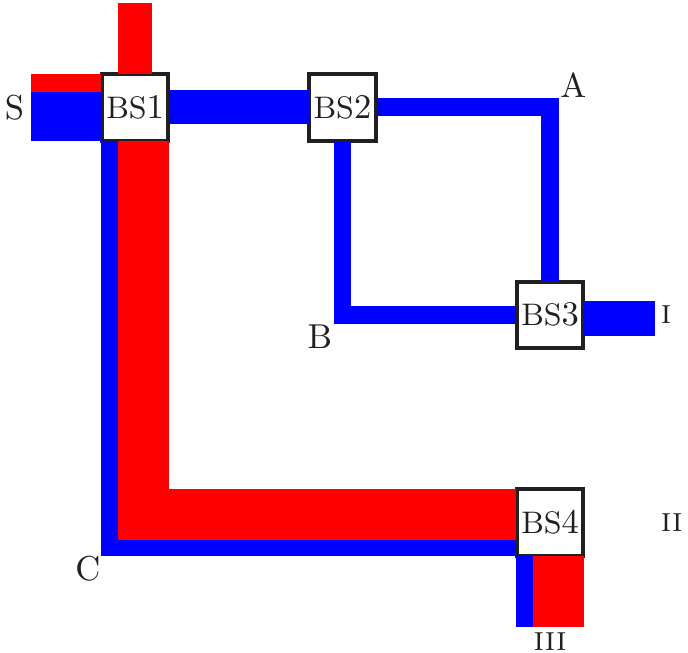}
  \caption{\label{fig:reply2}%
   The analog of Fig.~\ref{fig:reply1} when beam splitter BS4 has unit
   reflection probability.}
\end{figure}

As noted above, Fig.~\ref{fig:reply1} applies when the inner-loop
interferometer is balanced and ${0<R<1}$ with the
necessary adjustments of the line thickness.
For all intermediate $R$ values, then, we have Vaidman's narrative that a
particle detected at exit \textsc{iii} has earlier left traces at checkpoint A
and at checkpoint B and also at checkpoint C.
For ${R=1}$, we have Fig.~\ref{fig:reply2} with vanishing weak values at
checkpoints A and B.
In this situation, Vaidman's narrative is that a particle detected at exit
\textsc{iii} has earlier left a trace at checkpoint C, but did not leave
traces at checkpoints A and B.
Accordingly, a last-moment choice between ${R=\frac{1}{3}}$ and ${R=1}$ is a
choice between these two different pasts of the quantum particle, at a time
when the trace-leaving (or not) has already happened earlier.
While retrocausation of this kind is disquieting for Vaidman's
definition of the particle's past, it has no bearing on the story told by
standard quantum mechanics: 
Both in Fig.~\ref{fig:reply1} and in Fig.~\ref{fig:reply2}, the particle
passes through checkpoint C only on its way from the source S to the
exit~\textsc{iii}.


\begin{thebibliography}{1}

\bibitem{Peleg+1:18}
U. Peleg and L. Vaidman,
\textit{Comment on ``Past of a quantum particle revisited''},
\href{https://arxiv.org/abs/1805.12171v1}{e-print arXiv:1805.12171v1}
(to appear in \pra).

\bibitem{Englert+4:17}
B.-G. Englert, K. Horia, J. Dai, Y. L. Len, and H.~K.~Ng,
\textit{Past of a quantum particle revisited},
\pra\ \textbf{96}, 022126 (2017);
arXiv version:
\textit{Past of a quantum particle: Common sense prevails},
\href{https://arxiv.org/abs/1704.03722v3}{e-print arXiv:1704.03722v3}.

\bibitem{Englert+3:92}
B.-G. Englert, M. O. Scully, G. S\"ussmann, and H. Walther,
\textit{Surrealistic Bohm Trajectories},
Z. Naturforsch.\ \textbf{47a}, 1175 (1992).

\bibitem{Aharonov+1:96}
Y. Aharonov and L. Vaidman,
\textit{About Position Measurements Which Do Not Show the Bohmian Particle
  Position}, in \textit{Bohmian mechanics and quantum theory: An appraisal},
edited by T. Cushing, A. Fine, and S. Goldstein,
Boston Studies in the Philosophy of Science, Vol.~184 
(Kluwer, 1996), pp.~141--154.

\bibitem{fn1}
A detailed treatment can be found in Sec.~IV of \cite{Englert+4:17}.

\bibitem{fn2}
The suggestion in \cite{Peleg+1:18} that a particle is ``everywhere where the
wave function is non vanishing'' is equally at odds with standard quantum
mechanics.

\bibitem{fn3}
Since there is, in fact, no such equation in \cite{Peleg+1:18}, we provide it
here. 
  
\end{thebibliography}
\end{document}